\documentstyle[preprint,aps,epsf]{revtex}
\tightenlines
\begin{document}
\draft
\preprint{quant-ph/9708037}
\title{Moments of the Wigner Distribution and a Generalized
Uncertainty Principle} 

\author{R. Simon$^{1}$ and N. Mukunda$^{2}$}

\address{$^1$The Institute of Mathematical Sciences, C.I.T. 
Campus, Madras 600 113, India\\
$^2$Centre for Theoretical Studies, Indian Institute of Science,
Bangalore 560 012, India ~~and  \\
Jawaharlal Nehru Centre for Advanced Scientific Research,
Jakkur, Bangalore 560 064, India}

\date{\today}
\maketitle
\begin{abstract}
The nonnegativity of the density operator of a state is faithfully coded
in its Wigner distribution, and this places constraints on the moments of the
Wigner distribution. These constraints are presented in a canonically 
invariant form which is both concise and explicit. 
Since the conventional uncertainty principle is such a constraint on the
first and second moments, our result constitutes a 
generalization of the same to all orders. Possible
application in quantum state reconstruction using optical homodyne 
tomography  is noted.
\end{abstract}

\narrowtext

The uncertainty principle exhibits a fundamental manner in
which the quantum description of nature departs from the classical one.
For the canonical pair of variables $(\hat{q} , \hat{p})$ the Heisenberg
commutation relation $\left[ \hat{q}\,, \hat{p} \right] = {\rm i} \hbar$
leads, for any state $\left| \psi \right. \rangle$, to the unbeatable
limitation
\begin{equation}
\left\langle \left( \Delta \hat{q} \right) ^2 \right\rangle 
\left\langle \left( \Delta \hat{p} \right) ^2 \right\rangle 
-
\left\langle \frac{\Delta \hat{q} \Delta \hat{p} 
+ 
\Delta \hat{p} \Delta \hat{q} }{2} \right\rangle ^2
\ge \frac{\hbar ^2}{4}\,, 
\label{UNCERTAINTY}
\end{equation}
where $\langle \hat{q}\rangle$ 
$= \langle \psi \left| \hat{q} \right| \psi \rangle$,
$\Delta \hat{q} = \hat{q} - \langle \hat{q} \rangle$, and so on. 
Every Gaussian pure state saturates this inequality. An
important attribute of the uncertainty principle~(\ref{UNCERTAINTY}) is
that it is invariant under all real linear canonical transformations,
just as the canonical commutation relation is.

This inequality can be generalized in a naive manner to higher orders in 
$\hat{q}$, $\hat{p}$. For any
pair of hermitian operators $\hat{A}$, $\hat{B}$ and state 
$\left| \psi \right. \rangle$ we have the Schwartz inequality
\begin{equation}
\left\langle \hat{A} ^2 \right\rangle \left\langle \hat{B} ^2 
\right\rangle \ge
\left\langle \frac{\hat{A} \hat{B} + \hat{B} \hat{A} }{2} 
\right\rangle ^2 +
\left\langle \frac{\hat{A} \hat{B} - \hat{B} \hat{A} }{2{\rm i}} 
\right\rangle ^2\,. \label{SCHWARTZ}
\end{equation}
It is saturated if an only if $\hat{A} \left| \psi \right. \rangle$ and 
$\hat{B} \left| \psi \right. \rangle$ are linearly dependent as vectors.
Clearly,~(\ref{UNCERTAINTY}) is a particular case of~(\ref{SCHWARTZ})
corresponding to
$\hat{A} = \hat{q} - \langle \hat{q} \rangle$,
$\hat{B} = \hat{p} - \langle \hat{p} \rangle$. Clearly, the choice
$\hat{A} = \hat{q} ^2 - \langle \hat{q} ^2 \rangle$,
$\hat{B} = \hat{p} ^2 - \langle \hat{p} ^2 \rangle$
will lead to a higher order uncertainty principle involving
$\langle \hat{q} ^4 \rangle$, $\langle \hat{p} ^4 \rangle$;
the Fock states $\left| n \right. \rangle$, being eigenstates of
$\hat{q} ^2 + \hat{p} ^2$, will be expected to saturate this 
higher order uncertainty
principle. That they indeed do so can be explicitly verified.

One may indeed produce any number of such naive generalized
uncertainty principles by making various choices for $\hat{A}$,
$\hat{B}$ in~(\ref{SCHWARTZ}). But every one of them will suffer from
the deficiency of not being invariant under linear canonical 
transformations. Further, there seems to be no reasonable sense in which
the set of all such generalizations based on~(\ref{SCHWARTZ}) can be
considered to be complete.

The purpose of this Letter is to present a generalization of the
uncertainty principle which largely overcomes these difficulties. This is 
achieved by applying to the Wigner quasiprobability~\cite{Wigner} 
concepts and results from the
classical problem of moments\cite{Shohat}. The final result is a nested 
sequence of constraints on the moments
of the Wigner distribution. These constraints are tailored to capture the 
positivity of the density operator of a quantum state. Equivalently, a 
given real phase space distribution has to necessarily meet these constraints 
in order to qualify to be a bonafide Wigner distribution. 

It should be
appreciated that the higher moments of the Wigner distribution are no 
more objects of purely academic interest. An enormous 
progress in quantum state reconstruction 
using optical homodyde tomography has been achieved in the last few 
years: the
Wigner distribution of a state can now be fully mapped out~\cite{Vogel},
as has been demonstrated by several 
groups~\cite{Smithey,Breitenbach,Dunn,Leibfried}.

There exist rigorous and mathematically sophisticated approaches to the
quantum mechanical moment problem~\cite{Kastler}. But our  
considerations here are explicit and take full advantage of the 
canonical invariance underlying the Heisenberg commutation relation.

Details of a classical probability density $\rho ( x )$ are coded in its
moments $\gamma _n = \int d x x^{n} \rho ( x )$. An
important result in the problem of moments is this~\cite{Shohat}: given
a sequence of numbers it qualifies to be the moment sequence of a
bonafide probability distribution if and only if the symmetric matrix
defined below is nonnegative:
\begin{equation}
\Gamma =  
\left(
\begin{array}{cccc}
\gamma _0 & \gamma _1 & \gamma _2 & \cdots \\
\gamma _1 & \gamma _2 & \gamma _3 & \cdots \\
\gamma _2 & \gamma _3 & \gamma _4 & \cdots \\
\vdots    & \vdots    & \vdots    &  \cdots \\
\end{array}
\right) \ge 0\,.
\label{GAMMA}
\end{equation}
This can be broken into a sequence of positivity conditions on the 
determinants of
the submatrices of $\Gamma$, which in turn can be viewed as a nested 
sequence of constraints on the moments $\gamma _n$; and these constraints
are tailored to capture the pointwise nonnegativity of $\rho ( x )$. 
Reconstruction of $\rho ( x )$ from its moment sequence is the
other part of the classical problem of moments~\cite{Shohat}.

In quantum mechanics, the state is described not by a true probability
density in phase space, but by one of several possible
quasiprobabilities~\cite{Wigner}. The earliest, and probably the most 
prominent,
quasiprobability is the one introduced by Wigner~\cite{Wigner}. It is
intimately related to the Weyl ordering rule of association between the
algebra ${\cal A}$ of functions $f \left( q , p \right)$  of the phase
space variables and the algebra $\hat{\cal A}$ of operator valued 
functions $\hat{F} \left( \hat{q} , \hat{p} \right)$ of the canonical
operators. The rule is specified first through the one to one
correspondence $e ^{\theta q + \tau p} \longleftrightarrow
e ^{\theta \hat{q} + \tau \hat{p}}$ for {\it plane waves}, and then extended
linearly to the entire algebra using Fourier techniques.

The Weyl rule could equally well be specified in the monomial basis
instead of the plane wave basis through the  association
$q^m p^n \longleftrightarrow
\hat{T} _{m , n}$ for $m$, $n = 0$, $1$, $2, \cdots$
where the Weyl ordered monomial $\hat{T} _{m , n}$ is the coefficient
of $\left( m ! n ! \right) ^{- 1} \theta ^m \tau ^n$ in the Taylor
expansion of $e ^{\theta \hat{q} + \tau \hat{p}}$. This is an
isomorphism 
between ${\cal A}$ and $\hat{\cal A}$ only at the level of vector spaces 
but not at the level of
algebras. In particular, the product of two $\hat{T} _{m , n}$'s is not
another monomial but a linear combination of 
monomials~\cite{Bender}:
\begin{eqnarray}
\hat{T} _{m , n} \hat{T} _{m ' , n '}
& = & 
\sum _{r , s} d _{r , s} \, 
\hat{T} _{ m + m ' - r -s\, , n + n ' - r - s}, \nonumber \\
d _{r , s} & = & 
\frac{ ( - 1 ) ^r {\left(\frac{{\rm i} \hbar}{2} \right)} ^{s + r}\,
m !\, n !\,}
{
\left(m - s \right) !\,
\left(n - r \right) !\, }
\left(
\begin{array}{c}
m ' \\
r
\end{array}
\right)
\left(
\begin{array}{c}
n ' \\
s
\end{array}
\right)\,. 
\label{MONOMIAL}
\end{eqnarray}

The intimate connection between Weyl ordering and Wigner distribution
is this: 
\begin{equation}
{\rm tr} ( \hat{\rho} \hat{T} _{m , n} )
= 
\int d q\, d p\, q ^m p ^n W \left( q , p \right)\,.
\label{WEIGHT}
\end{equation}
That is, the quantum mechanical expectation of the Weyl ordered monomial
$\hat{T} _{m , n}$ is precisely the $m n$-th moment of the Wigner
function. By linearity, similar relation holds for any pair
$f \left( q , p \right)$, 
$\hat{F} \left( \hat{q} , \hat{p} \right)$ related by Weyl ordering.

The monomials $\hat{T} _{m , n}$ are hermitian, and transform in a simple 
manner under 
the group ${\rm Sp}\left( 2 , \Re \right)$  of real linear canonical
transformations. This group can be identified with
${\rm SL}\left( 2 , \Re \right)$, the group of $2 \times 2$ real
matrices with unit determinant. ${\rm Sp}\left( 2 , \Re \right)$ acts
identically on the pairs $\left(q , p \right)$ and 
$\left(\hat{q} , \hat{p} \right)$, and this action induces linear 
transformation in the algebras ${\cal A}$ and $\hat{\cal A}$ in the 
natural manner.

The set of homogeneous polynomials of order $2j$ in $q$ and $p$~(being
linear combinations of $q ^{j -s} p ^{j + s}$ for $s = - j$,
$- j + 1\,, \cdots\,, j$~) transform linearly among themselves under this 
transformation, leading to the spin-$j$ representation
of ${\rm Sp}\left( 2 , \Re \right)$ in ${\cal A}$. The
$\hat{T} _{m , n}$'s in $\hat{\cal A}$ transform in the same manner as
the $q ^m p ^n$'s in ${\cal A}$, and thus the vector space
$\hat{\cal A}$ decouples into a direct sum of invariant subspaces
under ${\rm Sp}\left( 2 , \Re \right)$:
$
\hat{\cal A} = \hat{V} ^{( 0 )} \oplus\,
\hat{V} ^{( \frac{1}{2} )} \oplus\, \hat{V} ^{( 1 )} \oplus\, \cdots$
Clearly, $\hat{V} ^{( j )}$ is of dimension $2 j + 1$, and is spanned by
$\hat{\xi} _{j s} = \hat{T} _{j - s , j + s}$ with $s$ running over the
range $s = - j$, $- j + 1\,, \cdots\,, j$. It acts as the carrier space
for the spin-$j$ representation of ${\rm Sp}\left( 2 , \Re \right)$ in
$\hat{\cal A}$. Thus, every spin-$j$ representation of
${\rm Sp}\left( 2 , \Re \right)$ occurs in $\hat{\cal A}$ once and only
once.

It is convenient to arrange the $\hat{\xi} _{j , s}$'s for fixed $j$
into a $2j + 1$ dimensional 
column vector $\hat{\mbox{\boldmath $\xi$}} ^{( j )}$ and then,
for any chosen $J$, arrange these columns into a grand column vector
$\hat{\mbox{\boldmath $\xi$}} _J$ of dimension $(J + 1)(2J + 1)$.

Let the $\left( 2 j + 1 \right)\,\times\,\left( 2 j + 1 \right)$ matrix 
$K ^{( j )} \left( S \right)$  denote the spin-$j$ representation for
$S \in {\rm Sp}\left( 2 , \Re \right)$. Since the defining 
representation of ${\rm Sp}\left( 2 , \Re \right)$ is the
spin-$\frac{1}{2}$ representation, we have
$K ^{(\frac{1}{2})} \left( S \right) = S$. Let
$K _J \left( S \right)$ be the block diagonal matrix of order 
$\left(J + 1 \right)\,\times\,\left( 2 J + 1 \right)$ with diagonal
blocks $K ^{( 0 )} \left( S \right) = 1$,
$K ^{(\frac{1}{2})} \left( S \right)$,
$\cdots$, $K ^{( J )} \left( S \right)$. Then the action of
${\rm Sp}\left( 2 , \Re \right)$ in $\hat{\cal A}$ has the concise
description
\begin{eqnarray}
\hat{\mbox{\boldmath $\xi$}} _J \longrightarrow
K _J ( S ) \hat{\mbox{\boldmath $\xi$}} _J\,, \quad \quad
\hat{\mbox{\boldmath $\xi$}} ^{( j )} =
K ^{( j )} ( S ) \hat{\mbox{\boldmath $\xi$}} ^{( j )}\,.
\label{COMPACT}
\end{eqnarray}

We are now in a position to present the generalized uncertainty
principle.  For each~$J = 0$, $\frac{1}{2}$, $1, \cdots$ form the square
matrix $\hat{\Omega} _J$, of order 
$\left( J + 1 \right) \left( 2 J + 1 \right)$, with operator entries, 
through the definition~($\hat{\mbox{\boldmath $\xi$}} _J ^{\dagger}$
is a row vector with the same entries as the column vector
$\hat{\mbox{\boldmath $\xi$}} _J$)
\begin{eqnarray}
\hat{\Omega} _J =
\hat{\mbox{\boldmath $\xi$}} _J\,
\hat{\mbox{\boldmath $\xi$}} _J ^{\dagger}\,, \qquad
\left( \hat{\Omega} _J \right) _{j s , j' s'}
=
\hat{\xi} _{j s} \hat{\xi} _{j' s'}\,. 
\label{COLUMN}
\end{eqnarray}
We may write $\hat{\Omega} _J$ in more detail in the block form
\begin{eqnarray}
\hat{\Omega} _J =
\left(
\begin{array}{ccccc}
1 & \hat{\mbox{\boldmath $\xi$}} ^{( \frac{1}{2} ) ^{\dagger}} 
& \cdots
& \hat{\mbox{\boldmath $\xi$}} ^{( J ) ^{\dagger}}
\\
\hat{\mbox{\boldmath $\xi$}} ^{( \frac{1}{2} )} 
& \hat{\mbox{\boldmath $\xi$}} ^{( \frac{1}{2} )} 
  \hat{\mbox{\boldmath $\xi$}} ^{( \frac{1}{2} ) ^{\dagger}} 
& \cdots
& \hat{\mbox{\boldmath $\xi$}} ^{( \frac{1}{2} )} 
  \hat{\mbox{\boldmath $\xi$}} ^{( J ) ^{\dagger}}
\\
\vdots & \vdots & &  \vdots
\\
\hat{\mbox{\boldmath $\xi$}} ^{( J ) } 
& \hat{\mbox{\boldmath $\xi$}} ^{( J )} 
  \hat{\mbox{\boldmath $\xi$}} ^{( \frac{1}{2} ) ^{\dagger}} 
& \cdots
& \hat{\mbox{\boldmath $\xi$}} ^{( J )} 
  \hat{\mbox{\boldmath $\xi$}} ^{( J ) ^{\dagger}} \\
\end{array}
\right)\,. \nonumber
\end{eqnarray}
It is to be understood that each element of $\hat{\Omega} _J$ is written
as a linear combination of the $\hat{T} _{m , n}$'s
using~(\ref{MONOMIAL}). For purpose of illustration, we detail one of 
these blocks:
\begin{eqnarray}
\hat{\mbox{\boldmath $\xi$}} ^{( 1 )}\,
\hat{\mbox{\boldmath $\xi$}} ^{( \frac{1}{2} ) ^{\dagger}} =
\left(
\begin{array}{cc}
\hat{T} _{3 , 0} 
& \quad  \hat{T} _{2 , 1} + {\rm i} \hbar \hat{T} _{1 , 0} \\
\\
\hat{T} _{2 , 1} - \frac{{\rm i} \hbar}{2} \hat{T} _{1 , 0}
& \quad \hat{T} _{1 , 2} + \frac{{\rm i} \hbar}{2} \hat{T} _{0 , 1} \\
\\
\hat{T} _{1 , 2} - {\rm i} \hbar \hat{T} _{0 , 1} 
& \quad \hat{T} _{0 , 3} \\
\end{array}
\right)\,. \nonumber
\end{eqnarray}

Let $M _J = \langle \hat{\Omega} \rangle$ be the hermitian $c$-number
matrix obtained from $\hat{\Omega} _J$ by taking~(entrywise) quantum
mechanical expectation value in the given state $\hat{\rho}$:
\begin{eqnarray}
M _J
& = & 
{\rm tr} ( \rho\, \hat{\Omega} _J )\,
= \langle 
\hat{\mbox{\boldmath $\xi$}} _J
\hat{\mbox{\boldmath $\xi$}} _J ^{\dagger}
\rangle\,; \nonumber \\
\left( M _ J \right) _{j s , j' s'} 
& = &
{\rm tr} ( \hat{\rho}\, \hat{\xi}  _{j s}\, 
\hat{\xi} _{j' s'} ) 
\label{TRACE}
\end{eqnarray}
It will prove useful to write  $M _J$ in the block form
\begin{eqnarray}
M _J =
\left(
\begin{array}{ccccc}
1 & M ^{0 , \frac{1}{2}} & \cdots & M ^{0 , J} 
\\
M ^{\frac{1}{2} , 0} & M ^{\frac{1}{2} , \frac{1}{2}}
& \cdots & M ^{\frac{1}{2} , J}
\\
\vdots & \vdots & & \vdots \\
M ^{J , 0} & M ^{J , \frac{1}{2}} & \cdots & M ^{J , J} \\
\end{array}
\right)\,,
\label{BLOCK}
\end{eqnarray}
where $M ^{j , j'}$
$ = \langle \hat{\mbox{\boldmath $\xi$}} ^{\left( j \right)}
\hat{\mbox{\boldmath $\xi$}} ^{\left( j ' \right) ^{\dagger}} \rangle$
is a $\left( 2 j + 1 \right)\,\times\,\left( 2 j ' + 1\right)$
dimensional block, and
$M ^{j ' , j}$ $= \left( M ^{j , j'} \right) ^{\dagger}$. Since 
$\hat{\mbox{\boldmath $\xi$}} ^{\left( 0 \right)} = 1$,  $M 
^{0 , 0} = 1$ for all states. For purpose of illustration, we write out
a few leading blocks of $M _J$ explicitly: it is clear that the
row vectors
$M ^{0 , \frac{1}{2}}$ and $M ^{0 , 1}$ have entries 
$\left( \overline{q} , \overline{p} \right)$ and
$( \overline{q ^2} , \overline{q p} , \overline{p ^2} )$
respectively; further
\begin{eqnarray}
M ^{\frac{1}{2} , \frac{1}{2}}
=
\left(
\begin{array}{cc}
\overline{q ^2} & \quad \overline{q p} + \frac{{\rm i} \hbar}{2} \\
\\
\overline{q p} - \frac{{\rm i} \hbar}{2} & \quad  \overline{p ^2} \\
\end{array}
\right)\,, \nonumber 
\end{eqnarray}
\begin{eqnarray}
M ^{1 , \frac{1}{2}}
= 
\left(
\begin{array}{cc}
\overline{q ^3} 
& \overline{q ^2 p} + {\rm i} \hbar \overline{q}  \\
& \\
\overline{q ^2 p} - \frac{{\rm i} \hbar}{2} \overline{q}
& \overline{q p ^2} + \frac{{\rm i} \hbar}{2} \overline{p} \\
& \\
\overline{q p^2} - {\rm i} \hbar \overline{p} & \overline{p ^3} \\
\end{array}
\right)\,; \nonumber 
\end{eqnarray}
and finally, the $3 \times 3$ hermitian block $M ^{1 ,1}$ has the 
form
\begin{eqnarray}
\left(
\begin{array}{ccc}
\overline{q ^4}
& \overline{q ^3 p} + {\rm i} \hbar \overline{q ^2}  
& \overline{q ^2 p ^2} + 2 {\rm i} \hbar \overline{q p}
- \frac{\hbar ^2}{2} \\
& & \\
\overline{q ^3 p} - {\rm i} \hbar \overline{q ^2}
& \overline{q ^2 p ^2} + \frac{\hbar ^2}{4}
& \overline{q p ^3} + {\rm i} \hbar \overline{p ^2} \\
& & \\
\overline{q ^2 p ^2} - 2 {\rm i} \hbar \overline{q p}
- \frac{\hbar ^2}{2} 
& \overline{q p ^3} - {\rm i} \hbar \overline{p ^2}
& \overline{q ^4} \\
\end{array}
\right)\,. \nonumber
\end{eqnarray}
Here, $\overline{q ^m p ^n} = \langle \hat{T} _{m , n} \rangle$ stands
for the average of $q^m p ^n$ with the Wigner distribution as the weight
as in~(\ref{WEIGHT}).  In other words, $M _J$ is the {\it matrix formed out
of the  moments of the Wigner distribution function, of order atmost 
$2J$}.

We now prove the important fact that the nonnegativity of the density
operator $\hat{\rho}$ forces the hermitian matrix $M _J$ to be a
nonnegative matrix, for every $J$. For a given fixed value of $J$
consider the operator
\begin{eqnarray}
\hat{\eta} =
\sum _{j = 0} ^{J} \,
\sum _{s = - j} ^{j}\,
c _{j s}\, \hat{\xi} _{j s}\,, \nonumber
\end{eqnarray}
where $c _{j s}\,$ are arbitrary $c$-number expansion coefficients which
can be arranged into a
$\left( J + 1 \right) \left( 2 J + 1 \right)$ dimensional column vector
$C$. Now form the operator
\begin{eqnarray}
\hat{\zeta} = 
\hat{\eta} ^{\dagger}\, \hat{\eta}\, =
\sum _{j , s} \, \sum _{j' , s'}\, 
c _{j s} ^{*}\, c _{j' s'}\, \hat{\xi} _{j s}\, \hat{\xi} _{j' s'}\,,
\label{OPERATOR}
\end{eqnarray}
which is hermitian nonnegative by construction. Since
$\hat{\rho} \ge 0$, we necessarily have 
${\rm tr} ( \hat{\rho}\, \hat{\zeta} ) \ge 0$, for every
choice of the coefficients $\{ c _{j , s} \}$. But from~(\ref{TRACE}),
(\ref{OPERATOR}) we find
\begin{eqnarray} 
{\rm tr} ( \hat{\rho}\, \hat{\zeta} )\, & = &
\sum _{j , s} \, \sum _{j' , s'}\, 
c _{j s} ^{*}\, c _{j' s'}\, M _{j s , j' s'}\,.
\nonumber
\end{eqnarray}
That is, ${\rm tr} (\hat{\rho}\, \hat{\zeta} )$ $=$ 
$C ^{\dagger} M C$ for every $C$. This completes the proof that
$\hat{\rho} \ge 0$ implies $M _J \ge 0$ for every $J$.

A little reflection should convince the reader that this is the
generalized form of the uncertainty principle we have been after, and
we state it as follows:

\noindent
{\it Generalized Uncertainty Principle:}-- Let $M _J$ be the hermitian 
$c$-number matrix formed out of the moments of the Wigner distribution
of a state $\hat{\rho}$ in accordance with
the prescription~(\ref{TRACE}). Then
\begin{eqnarray}
M _J \ge 0\,, \qquad
J = 0\,, \frac{1}{2}\,,
1\,, \cdots\,
\label{M}
\end{eqnarray}
For a given state not all moments
will exist in general. It is clear that in such a case where $M _J$ is 
finite 
only for all $J \le J _{\rm max}$, our generalized uncertainty principle 
should be modified to
read $M _J \ge 0$, $J = 0$, $\frac{1}{2}$, $\cdots$,
$J _{\rm max}$.

While the hermiticity and unit trace properties of $\hat{\rho}$ are
reflected in the reality and normalization of the Wigner distribution, 
the generalized uncertainty principle presented in the concise matrix
form~(\ref{M}) exhibits the constraints on the moments
$\overline{q ^m p ^n}$ of the Wigner distribution
resulting from the nonnegativity of
$\hat{\rho}$.    
While the conventional uncertainty 
principle is such a constraint on the first and second moments, ours is a
generalization {\it to all orders}. It should be appreciated that the 
canonical commutation 
relation enters $M_{J}$ in~(\ref{M}) through~(\ref{MONOMIAL}).

The following mathematical lemma is helpful in analyzing the
content of this generalized uncertainty principle: A hermitian matrix 
$Q$ of the block form
\begin{eqnarray}
Q =
\left(
\begin{array}{cc}
A & C ^{\dagger} \\
C & B \\
\end{array}
\right) \nonumber
\end{eqnarray}
is positive definite if and only if $A$ and 
$B - C\, A ^{- 1}\, C ^{\dagger}$ are positive definite. 
The proof
simply consists in recognizing the congruence
\begin{eqnarray}
Q \sim Q' = L \, Q\, L ^{\dagger}\,, \quad \quad 
L =
\left(
\begin{array}{cc}
1 & 0 \\
- C\, A ^{- 1} & 1 \\
\end{array}
\right)\,, \nonumber
\end{eqnarray}
where $Q'$ is a block diagonal matrix with diagonal blocks $A$ and
$B - C A ^{- 1} C ^{\dagger}$.

The usual uncertainty principle (\ref{UNCERTAINTY}) is contained 
in~(\ref{M}) as a particular
case: it is equivalent to the condition
${\rm det} M _{J = \frac{1}{2}} \ge 0$.
Next consider the case $J = 1$. Use of the lemma with $C ^{\dagger}$
$= ( M ^{0 , \frac{1}{2}} \, M ^{0 , 1} )$ renders
$M _J \sim M _J '$, where 
\begin{eqnarray}
M _J '\, = 
\left(
\begin{array}{ccc}
1 & 0 & 0 \\ 
0 & M ^{\frac{1}{2} , \frac{1}{2}}
    - M ^{\frac{1}{2} , 0}\, M ^{0 , \frac{1}{2}} 
& M ^{\frac{1}{2} , 1} - M ^{\frac{1}{2} , 0}\, M ^{0 , 1} \\
0 & M ^{1 , \frac{1}{2}} - M ^{1 , 0}\, M ^{0 , \frac{1}{2}} 
  & M ^{1 , 1} - M ^{1 , 0}\, M ^{0 , 1} \\
\end{array}
\right)\,. \nonumber 
\end{eqnarray}
Now $M _J \ge 0$ implies $M _J ' \ge 0$ which in turn implies that its
diagonal block
$
M ^{\frac{1}{2} , \frac{1}{2}}
- M ^{\frac{1}{2} , 0} \, M ^{0 , \frac{1}{2}} \ge 0$.
Written in terms of the moments, the last condition reads
\begin{eqnarray}
\left(
\begin{array}{cc}
\overline{q ^2} - \overline{q} ^2  & \quad \overline{q p}
- \overline{q}\, \overline{p}
+ \frac{{\rm i} \hbar}{2} \\
& \\
\overline{q p} - \overline{q}\, \overline{p} 
- \frac{{\rm i} \hbar}{2} & \quad
\overline{p ^2} - \overline{p} ^2 \\
\end{array}
\right)\, \ge 0\,,
\label{CONDITION-1}
\end{eqnarray}
which is precisely the usual uncertainty 
principle~(\ref{UNCERTAINTY}).

One more application of the lemma on the nontrivial part of $M _J '$ 
further strengthens the positivity requirement on the other diagonal
block $M ^{1 , 1} - M ^{1 , 0}\, M ^{0 , 1}$ to  
\begin{eqnarray}
M ^{1 , 1} - M ^{1 , 0} M ^{0 , 1}
& \ge &  
C
\left(
M ^{\frac{1}{2} , \frac{1}{2}} - M ^{\frac{1}{2} , 0} 
M ^{0 , \frac{1}{2}} \right) ^{ - 1}
C ^{\dagger}\,, \nonumber \\
C  & = &
\left(
M ^{1 , \frac{1}{2}}
-
M ^{1 , 0}
M ^{0 , \frac{1}{2}} 
\right)\,. 
\label{CONDITION-2}
\end{eqnarray}
This $3 \times 3$ matrix condition, together with the $2 \times 2$
matrix condition~(\ref{CONDITION-1}), constitutes a complete statement
of the generalised uncertainty principle involving moments of all order 
upto and including the fourth.

It is clear that yet another application of the lemma, starting with
$M _{J = \frac{3}{2}}$, will lead to a positivity statement on a
$4 \times 4$ matrix which, together with~(\ref{CONDITION-1}) 
and~(\ref{CONDITION-2}), will constitute a complete statement of our
uncertainty principle on moments of all orders upto and including the
sixth. Evidently, this reduction algorithm based on the above lemma can
be continued to any desired value of $J$, and hence upto any
desired~(even) order of the moments,{\it eventually rendering $M _J$ block
diagonal}. 

We see from~(\ref{COMPACT}), (\ref{TRACE}) that $M _J$
transforms in the following manner under $S \in {\rm Sp} 
\left(2 , \Re \right)$:
\begin{eqnarray}
S: \quad \quad
M _J \longrightarrow K _J ( S )\, M _J\, K _J ( S ) ^T\,.
\label{CANONICAL}
\end{eqnarray}
The nonnegativity of $M _J$ is manifestly preserved under this 
transformation. Thus, our generalized uncertainty principle is invariant
under linear canonical transformations. Further, the 
reduction algorithm suggested by the lemma is invariant under linear 
canonical transformations, for it follows 
from~(\ref{COMPACT}) and~(\ref{CANONICAL}) that $M ^{j , j '}$ 
transforms to 
$K ^j ( S ) M ^{j , j '} K ^{j '}( S ) ^{T}$ under
$S \in {\rm Sp}\left( 2 , \Re \right)$.

An evidently useful way of reading~(\ref{CANONICAL}) is that the
components of $M _J$, just as the ${\hat{T} _{m , n}}$'s, transform as
tensors under ${\rm Sp}\left( 2 , \Re \right)$. And the fact that our
generalized uncertainty principle is invariant under 
${\rm Sp}\left( 2 , \Re \right)$ means that it is implicitly stated in
terms of the invariants of these tensors. These invariants, in the
classical case, have been studied in great detail by Dragt and 
coworkers~\cite{Dragt}.

While the nonnegativity of $\hat{\rho}$ implies the nonnegativity of
$M _J$ for all $J$, it is of interest to know if nonnegativity of $M _J$
for all $J$ implies nonnegativity of $\hat{\rho}$. Phrasing it somewhat
differently, we may ask: Given a real normalized phase space 
distribution whose moments satisfy the condition $M _J \ge 0$, for all
$J$, does it follow that the phase space distribution is a bonafide
Wigner distribution?

From the very construction of $M_J$, it is clear that 
${\rm tr} ( \hat{\rho} \hat{\cal O} ) \ge 0$ when
$\hat{\cal O}$ is of the form $\hat{\zeta}$ in~(\ref{OPERATOR}).
By linearity, this is true also when $\hat{\cal O}$ is a~(convex) 
linear combination of operators of this type(with nonnegative
coefficients). Thus,~(\ref{M}) will be sufficient to characterise the Wigner
distribution if the set of all such convex combinations is dense in the
space of nonnegative operators. Intuitively, this may appear to be the
case. However, the monomials $\hat{T} _{m , n}$ are generically noncompact,
and hence  a careful analysis of the issue of convergence should be made
before one can make any claim in this direction.

We have already referred to the intensity with which current
experimental research dealing with measurement of the Wigner 
distribution is being pursued~\cite{Smithey,Breitenbach,Dunn,Leibfried}. 
Since 
measurements are always accompanied by errors of various origins, it will 
be of interest to see to what extent the Wigner distribution reconstructed 
in a real experiment respects the generalised uncertainty principle. 
Further, 
it may be of interest to examine the possibility of incorporating these 
fundamental inequalities in the algorithm for tomographically reconstructing 
the 
Wigner distribution from measured data, in such a way as to improve the 
reconstruction itself.
Finally, our analysis applies equally well to any other quasiprobability, 
provided we choose suitably ordered momomials and modify the product 
formula~(\ref{MONOMIAL}) accordingly.

\end{document}